\documentclass[sigconf]{acmart}

\usepackage[utf8]{inputenc}
 
\usepackage{listings}
\usepackage{xcolor}
 
\definecolor{codegreen}{rgb}{0,0.6,0}
\definecolor{codegray}{rgb}{0.5,0.5,0.5}
\definecolor{codepurple}{rgb}{0.58,0,0.82}
\definecolor{backcolour}{rgb}{0.95,0.95,0.92}
 
\lstdefinestyle{mystyle}{
    backgroundcolor=\color{backcolour},   
    commentstyle=\color{codegreen},
    keywordstyle=\color{magenta},
    numberstyle=\tiny\color{codegray},
    stringstyle=\color{codepurple},
    basicstyle=\ttfamily\footnotesize,
    breakatwhitespace=false,         
    breaklines=true,                 
    captionpos=b,                    
    keepspaces=true,                 
    numbers=left,                    
    numbersep=5pt,                  
    showspaces=false,                
    showstringspaces=false,
    showtabs=false,                  
    tabsize=2
}

\AtBeginDocument{%
  \providecommand\BibTeX{{%
    \normalfont B\kern-0.5em{\scshape i\kern-0.25em b}\kern-0.8em\TeX}}}

\setcopyright{acmcopyright}
\copyrightyear{2018}
\acmYear{2018}
\acmDOI{10.1145/1122445.1122456}

\acmConference[PASC '20]{PACS '20: Platform for Advanced Scientific Computing}{June 29--July 01, 2020}{Geneva, Switzerland}
\acmBooktitle{PACS '20: Platform for Advanced Scientific Computing,
  June 29--July 01, 2020, Geneva, Switzerland}
\acmPrice{15.00}
\acmISBN{978-1-4503-9999-9/18/06}

\begin{document}

\title[Hardware locality-aware partitioning and dynamic load-balancing]{Hardware locality-aware partitioning and dynamic load-balancing of unstructured meshes for large-scale scientific applications}

\author{Pavanakumar Mohanamuraly}
\orcid{0000-0001-9788-8664}
\email{mpkumar@cerfacs.fr}
\affiliation{%
  \institution{CERFACS}
  \streetaddress{}
  \city{Toulouse}
  \state{France}
  \postcode{}
}

\author{Gabriel Staffelbach}
\email{gabriel.staffelbach@cerfacs.fr}
\affiliation{%
  \institution{CERFACS}
  \streetaddress{}
  \city{Toulouse}
  \state{France}
  \postcode{}
}

\renewcommand{\shortauthors}{Mohanamuraly et al.}

\begin{abstract}
  We present
  an open-source
  topology-aware
  hierarchical
  unstructured mesh
  partitioning and
  load-balancing tool
  \emph{TreePart}.
  The framework
  provides 
  powerful
  abstractions
  to automatically
  detect and build
  hierarchical
  MPI topology
  resembling the
  hardware at
  runtime. Using this
  information
  it intelligently
  chooses between
  shared and distributed
  parallel algorithms
  for partitioning
  and load-balancing.
  It provides a range
  of partitioning
  methods by interfacing
  with existing
  shared and distributed
  memory parallel
  partitioning
  libraries.
  It provides powerful 
  and scalable abstractions 
  like one-sided distributed 
  dictionaries and MPI3 shared
  memory based halo communicators
  for optimising HPC codes.
  The tool was
  successfully integrated
  into our in-house code
  and we present
  results from a
  large-eddy simulation
  of a combustion problem.
\end{abstract}





\maketitle

\section{Introduction}
The road to exascale has many 
facets, on the one end
are the challenges of hardware 
resilience and data storage
and on the
other is the challenge
of scaling existing 
scientific software
on exascale hardware.
Heterogeneous architectures
(accelerators) with
hybrid parallelisation
techniques so far have
shown good promise but
come with many
drawbacks. The first problem
is the shear abundance of
hybrid programming models.
Every HPC vendor seem to have
a solution:
Intel OneAPI~\cite{Silveira2000}, 
Kokkos~\cite{CarterEdwards2014}, 
RAJA~\cite{Hornung2015},
Intel TBB, 
Charm++~\cite{White2018},
CUDA,
Chapel~\cite{Chamberlain2007},
etc. Even
standards are many OpenMP,
OpenACC, OpenCL, OMPss, etc.
Most times features
from a latest version or standard
is necessary for better
performance but might
not be portable. Also
some of theses models
have a steep
learning curve to
write performant code.
This leads us to 
the second major problem
of portability (of both
platform and performance).
The third major problem
is the issue of
software maintenance
since the programming
models require major
disruption to the
code to achieve
best performance.

Even though one
has a variety of
choices for
the SMP
computing 
the choice is quite limited
for the off-processor model. In
fact MPI is the de facto
standard. The success
of MPI has been its
simplicity and portability.
Armed with just few MPI
calls it is possible to
run the code on both SMP
and off-processor nodes. 
But the MPI
implementations have remained
stagnant and have responded
only when challenged by
competing programming models.
In addition safely
mixing SMP models with
MPI is still problematic
with little or
no debugging support
for an application developer.
For example supporting
MPI calls inside threads/tasks
is still unsupported
in many vendor MPI libraries
and not widely
portable or available.

Nevertheless, with the advent
of the wide adoption of 
the MPI3 and
maturity of the library
ecosystem it is now
possible to mitigate
most of the performance
problems associated with 
a pure MPI implementation.
The Exascale MPI and OMPI-X
projects have already begun
to address the various
needs of exascale computing
and the MPI4 standard is
in the making. 
Rafenetti~\cite{Rafenetti2017} shows the various optimisations
undertaken in MPICH2 library
to mitigate issues with
large number of MPI process
or ranks. 
Balaji~\cite{Balaji2011,Thakur2010} et al.
discuss running MPI on
millions of cores and 
provide techniques to 
alleviate scalability limitations
in applications.

Our in-house HPC combustion
code AVBP\cite{schonfeld1999steady, Gourdain2009a,Gourdain2009b}
is a conglomerate effort
of several years of research
and development.
It has many stakeholders
ranging from
universities, research labs and
industrial partners. Therefore
we exercise extreme prudence
before bringing disruptive 
changes since every
change must be validated
across our value chain.
We investigated
the sources
of the issues
associated
with running MPI
on many ranks (typically 10k-15k).
We found satisfactory
solutions without resorting
to hybrid-programming
and with minimal disruption to
our code. To our
surprise every solution
revolved around one theme,
\textit{locality-awareness}.

We organise the paper as follows: (i) outline the problems associated with MPI for large number of ranks, (ii) present TreePart library and standalone tools which are aimed at solving most of the aforesaid problems, and (iii) substantiate with results from our in-house code AVBP (iv) followed by an outline on our future direction of research.

\section{Problems with MPI scaling}
Parallel unstructured mesh solvers
behave much differently when
compared to structure or
multi-block structured meshes.
The computation in the case
of unstructured meshes are
irregular causing problems
of indirect, non-strided, or
non-contiguous access to
memory. In addition they
require the use of more
sophisticated tools
to partition the mesh.
Structured mesh problems
are much easier to tune
and hence tuned library
implementations are
ubiquitous. This is hardly
true in the case of
unstructured meshes.
A summary of 
the most critical
performance problems
running parallel 
unstructured mesh
based solvers is shown
with relevant results.
% HPC developers often times
% brand problems that are of non-MPI
% in nature to problems of MPI.
% Hybrid parallelisation approaches
% alleviate some of these
% problems. But few of them
% still exist, although less severe,
% due to the lower number of partitions
% compared to a pure MPI
% implementation. 
% We present some of 
% the most important
% bottlenecks we faced
% while running our
% in-house solver
% on large number
% of MPI ranks
% and summarise them
% below.
% Some of these
% have also been pointed out
% in the Adaptive MPI library 
% (AMPI)~\cite{White2018,Rodrigues2010}
% based on Charm++.

\subsection{Graph partitioning at scale}
When the problem size (mesh size) becomes quite
large the memory requirements of serial graph 
partitioning no longer fit even on a fat-memory
node. Therefore, distributed graph libraries
like ParMETIS~\cite{Karypis1999}, PT-Scotch~\cite{Chevalier2008},
KaHip~\cite{Schulz2017}, 
Zoltan~\cite{Boman2012}, 
etc., are
employed for partitioning
the mesh. 
Mohanamuraly
et al.~\cite{Pavanakumar2013}
and Wang~\cite{Wang2012} et al.
show that
the quality of 
partitions 
(number of edge cuts)
obtained from
graph partitioning tools
deteriorates rapidly when the
number of partitions increases.
% and beyond a threshold
% the graph partitioner is
% only as good as a geometric
% partitioner.
In fact 
the partition quality
considering the same 
underlying mesh
using a parallel partitioner
is worser than the
serial
algorithm~\cite{Karypis1998}.

In an online partitioning approach
for unstructured meshes
we ideally like to have every
subscribed rank to participate
and construct its local mesh.
In our experience when we used
core counts in excess of $5k$
(also MPI ranks) most graph
partitioners fail due to the
presence of all-to-all type
communication in their implementation.
Therefore we resort
to an off-line partitioning strategy
which limits the scope for
online load-balancing and 
associated mesh partitioning.
KaHIP~\cite{Akhremtsev2015}
provides
semi-external and external
graph algorithms which
can be used to partition
very large meshes
on a single rank but
limits the scalability.
In hybrid programming models
the above problem is less
severe (but still present)
because the mesh is partitioned
for physical nodes leading
to lower partition counts.

\subsection{Partition-to-rank assignment}

\begin{figure}
    \centering
    \includegraphics[scale=0.4]{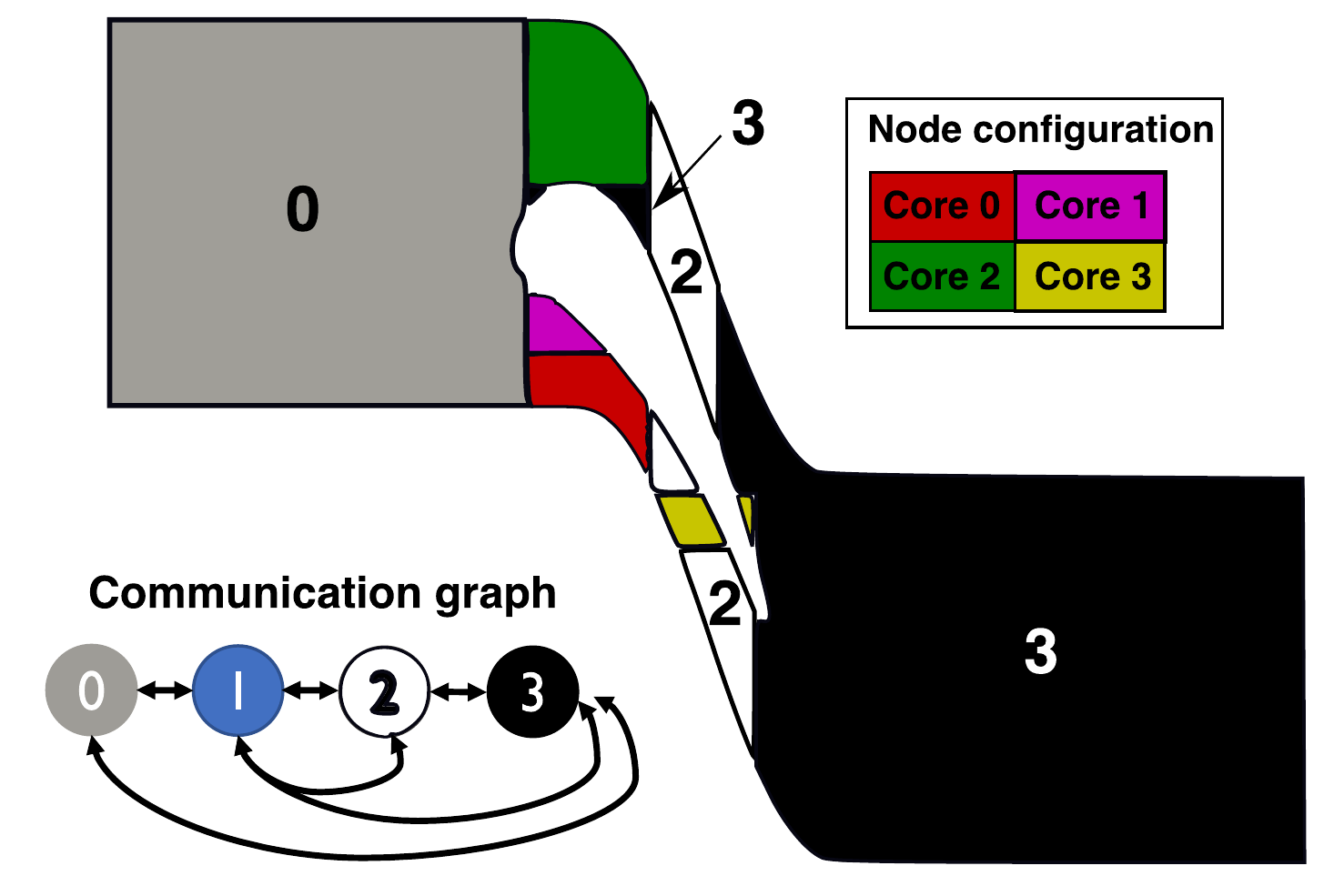}
    \caption{Sub-optimal partition to rank assignment}
    \label{fig:no-hier-part}
\end{figure}

\begin{figure}
    \centering
    \includegraphics[scale=0.4]{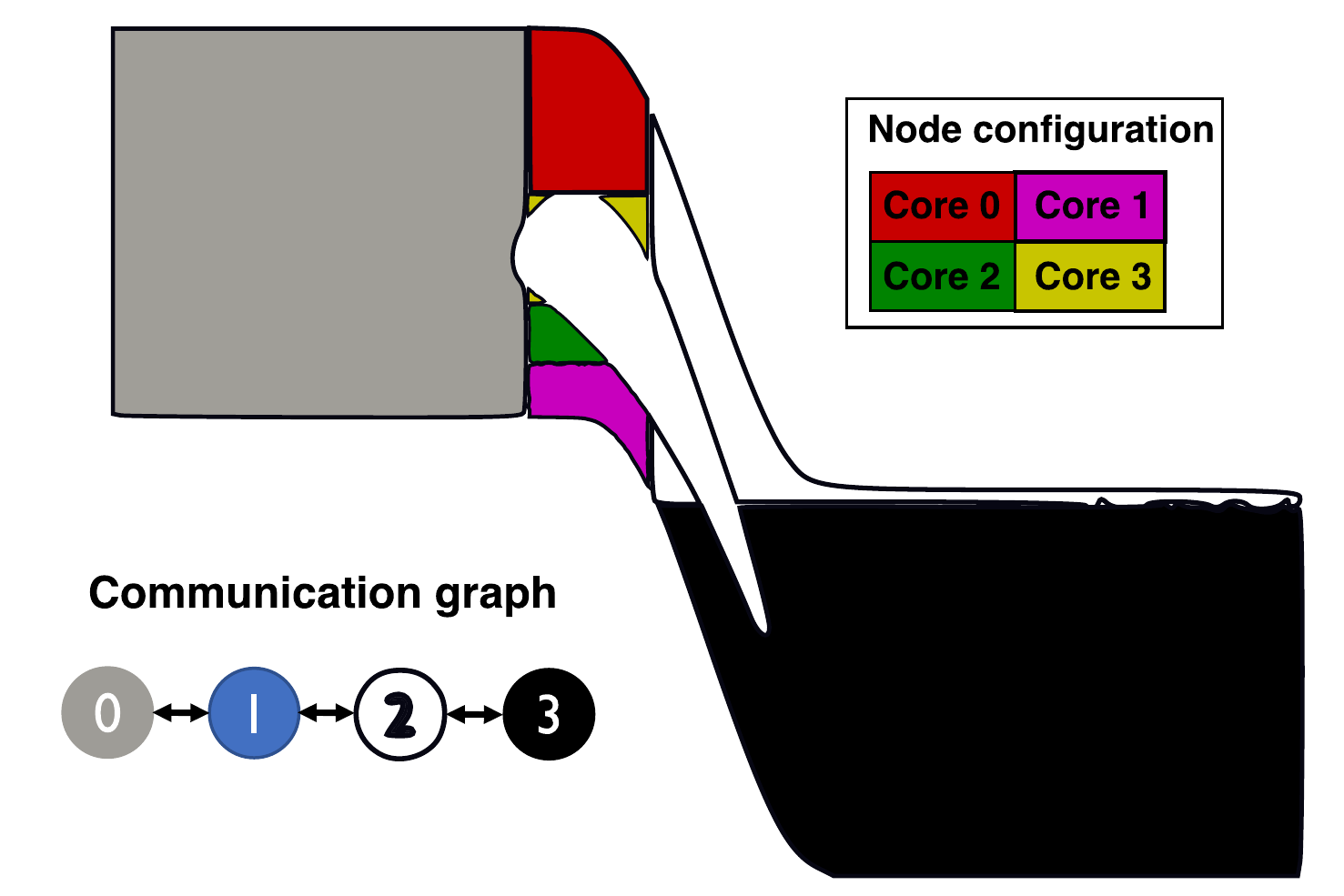}
    \caption{Optimal partition to rank assignment}
    \label{fig:hier-part}
\end{figure}

This is an often overlooked problem in
practical MPI implementations. Once a
quality partitioning is obtained
they have to be optimally placed
or pinned to the physical cores 
reserved for the computation.
Mapping each partition to MPI
ranks in the natural ordering
mostly results in poor performance.
In figure~\ref{fig:no-hier-part}
we show an unstructured
mesh whose partitions are
sub-optimally pinned to
four nodes (with four cores each).
Note that the core partitions
(coloured red, pink, yellow, and green)
in node 1 are scattered and 
share boundaries with many
node neighbours. We also show an
optimal node-to-rank placement
in figure~\ref{fig:hier-part}. 
MPI standard actually
allows such optimised placement
using MPI distributed
graph topology~\cite{Hoefler2009,Rashti2011}. Unfortunately
from our experience
we never encountered an MPI
library that actually does the
optimal placement for 
distributed graphs.
Similar observations were
made by 
Hoefler et al.~\cite{Hoefler2011}
who states that MPI mapping
function are often not
well implemented.
Graph partitioning
library
KaHIP~\cite{Schulz2017}
provides optimal 
processor mapping
of partitions
by solving a
sparse quadratic 
assignment problem.
Note that
hybrid programming models
avoid this issue completely
since the partitions are
always at the node level.


\subsection{Load imbalance}
\begin{figure}
    \centering
    \includegraphics[scale=0.5]{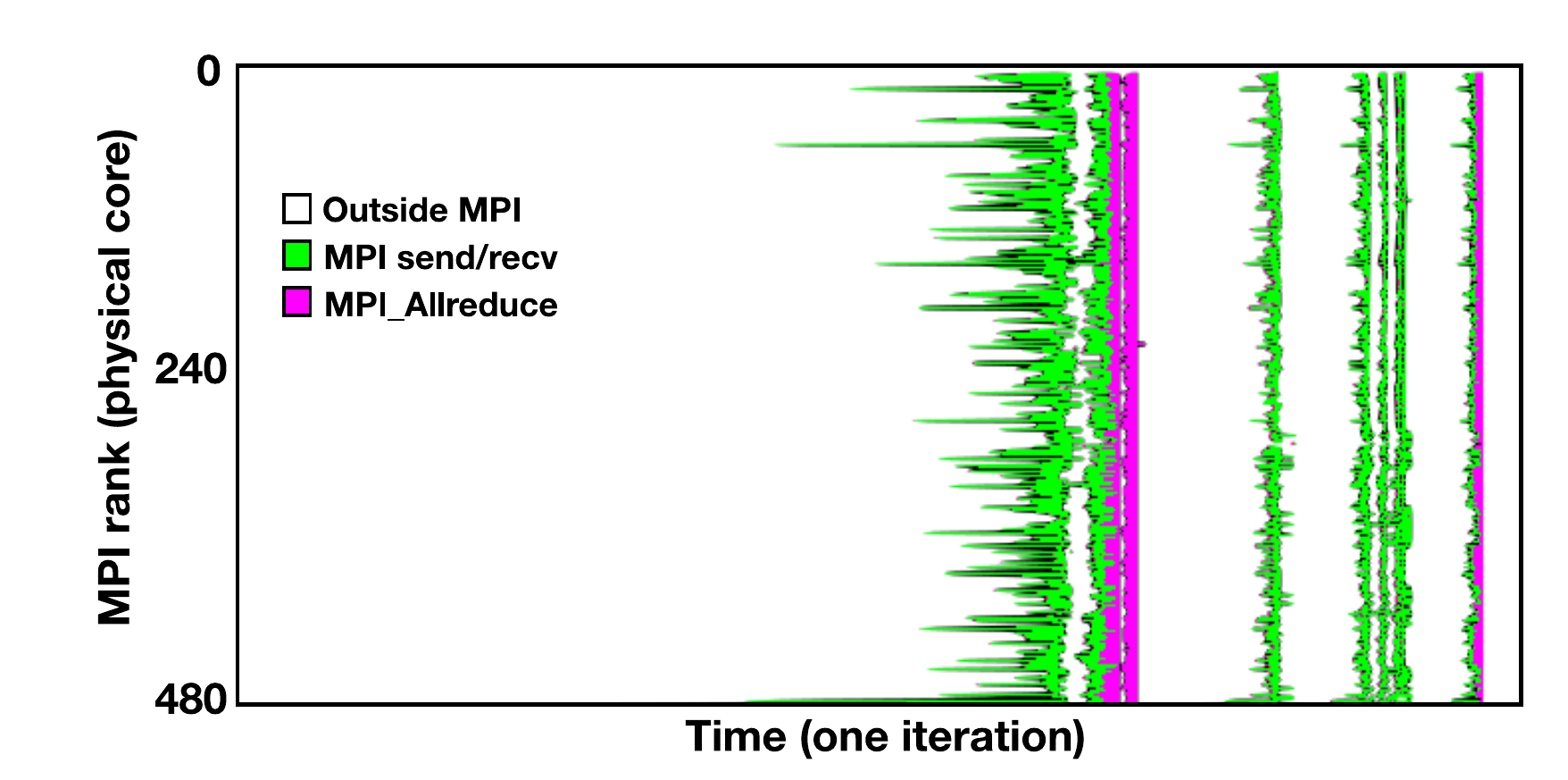}
    \caption{Load imbalance in one iteration of the solver for an unstructured mesh with 150M elements on 480 MPI ranks (physical cores)}
    \label{fig:imbal-graph-part}
\end{figure}

Load imbalance has been an important argument
in favour of leaning towards hybrid
programming compared to a pure MPI
implementation. The runtime libraries
like OpenMP, OpenACC, TBB, etc.
automatically load-balance
tasks/threads. But one needs
to specify the right grain size
for each task and the load-balancing
strategy to obtain optimal
performance. Usually the
graph partitioner generates
mesh partitions with an
imbalance in element count
within $1-2\%$. Certain
additional operations
like boundary conditions,
numerical kernels that
selectively execute based
on solution evolution,
contribute to additional
load-imbalance.
In figure~\ref{fig:imbal-graph-part}
we show the load imbalance
%measured 
in our in-house code
for an unstructured
mesh with 150M
tetrahedral
elements, partitioned 
using ParMETIS,
for different partition
sizes. We clearly
see a large imbalance
in load even though
the graph partitioner
has under 2\% vertex
imbalance.
Perfect load balancing is 
seldom possible in 
most CFD applications, 
especially when considering 
industrial configurations.
% Interestingly we see that the load imbalance
% in graph partitioner deteriorates rapidly
% with increase in number of partitions.
%Therefore a 
A pure MPI implementation
%is bound to 
can suffer performance
degradation since automatic 
load-balancing
is not available in most
MPI run time implementations
except for AMPI, where
the runtime calls
user implemented
packing/unpacking routines
for automatic migration of 
MPI process. Note that
even in a hybrid environment
node level imbalances
cannot be mitigated
as a result of this
issue.

The load-balancing of tasks
or threads in a hybrid model
is highly essential and
built into the run-time
implementation.
In the case of
a pure MPI 
implementation a load
imbalance would trigger a
re-partitioning of the mesh
or one needs to resort to the
idea of over-subscription
adopted by
AMPI~\cite{White2018}.
The over-subscribed case
suffers even more with
partition quality
because the number of
partitions are 
substantially
higher than the 
partition-per-core
MPI model.

Re-partitioning
unstructured
meshes to
improve load imbalance
requires the
use of weights
defined at mesh
entities like
nodes or elements.
The weights are
estimated using
coarse grained
measurements like
compute time
which need not
directly correlate to
the fine grain behaviour.
In the presence
of multiple computational
kernels each exhibiting
different or conflicting
load imbalance it is
impossible to
accurately capture
the behaviour
using a single weight.
A multi-constraint
partitioning~\cite{Karypis2014}
is
necessary to
effectively
handle
such a
situations.
% of the MPI process.
% Note that this problem has little bearing
% on the MPI model but more to do with
% the deficiencies of the partitioner.
\subsection{Intranode messaging}

The intranode messaging is one of the
most important performance optimisation
in a pure MPI implementation.
The three main
approaches to MPI
intranode communication are
network loopback, user-level
shared memory, and kernel 
assisted direct
copy~\cite{Chai2008}.
Network loop back has 
poor performance and
seldom used nowadays.
User-level shared memory
is more
widely implemented
in MPI libraries.
Kernel assisted
direct copy approaches 
provide superior performance
by eliminating intermediate
copies.
Linux kernel (v3.4 and above) has
native support for 
Cross Memory Attach (CMA)
to enable a single
rather than a
double copy of MPI messages
via shared memory.
CMA for intranode
communication is supported
by MPI libraries like
Intel MPI, OpenMPI, etc.
This is a less severe
bottleneck on newer
MPI implementations when
direct kernel copy
is supported by the
kernel.% but can become
%a bottleneck if this is
%not supported.
\begin{figure}
    \centering
    \includegraphics[scale=0.6]{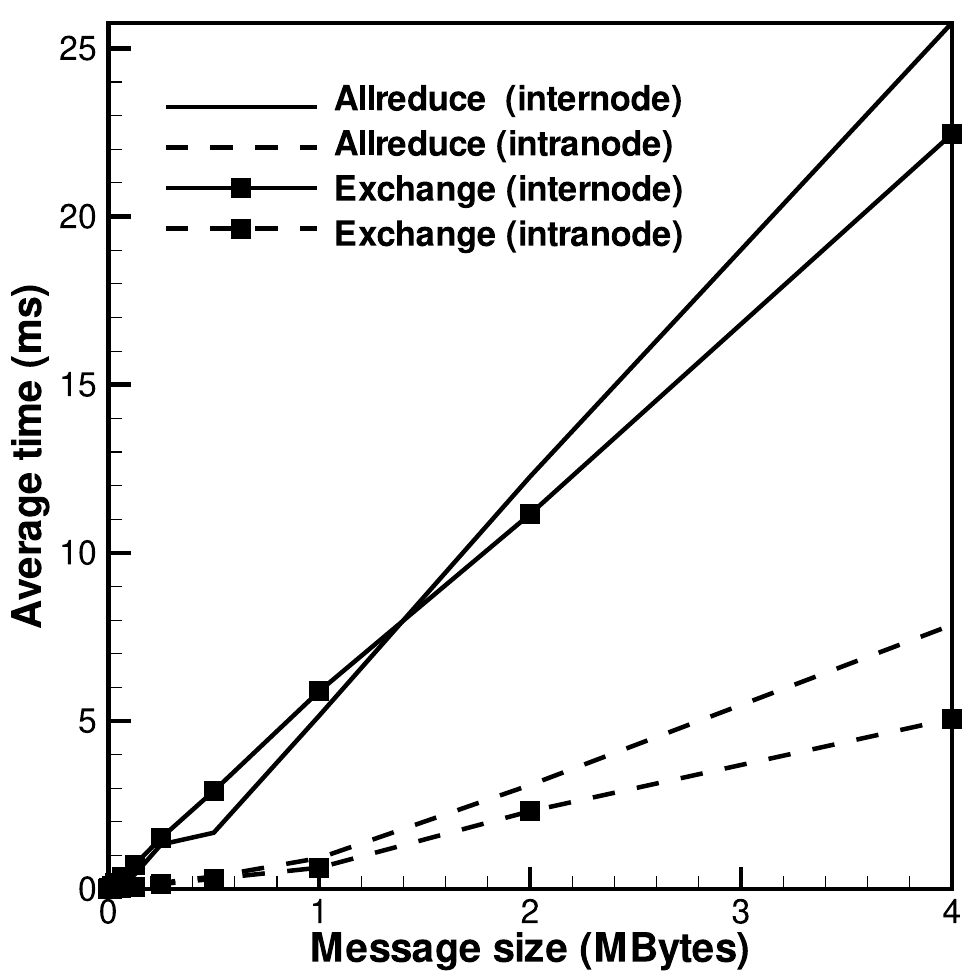}
    \caption{MPI all-reduce and exchange benchmark timing results for inter and intranode communication (Intel MPI 2018.1, IMB-MPI1 2019-u3 and Infiniband hardware)}
    \label{fig:mpi-ping-pong}
\end{figure}

\subsection{Communication imbalance}

\begin{figure}
    \centering
    \includegraphics[scale=0.7]{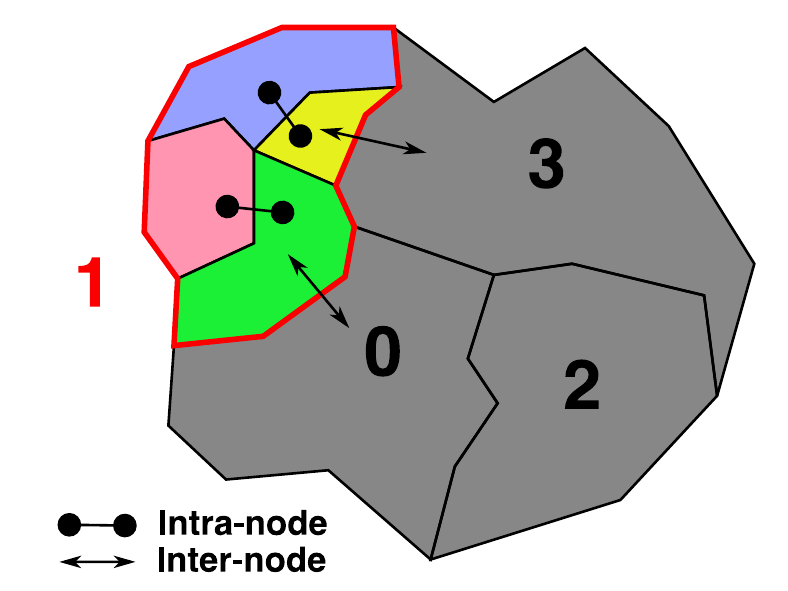}
    \caption{Communication imbalance in a partitioned mesh; $0-3$ are the node partitions and partitions within node $1$ are coloured to indicate the core partitions}
    \label{fig:comm-imbal}
\end{figure}

A major component
contributing to
imbalance in execution times
between MPI ranks
is communication
imbalance. 
In figure~\ref{fig:mpi-ping-pong}
the IMB-MPI1 benchmark timing
obtained for all-reduce and exchange
is shown for inter and intranode
placements. We see an
almost tripling of network bandwidth
in the case of intranode placement.
Therefore even with an optimal
placement of ranks to physical cores
a pure MPI implementation suffers
from
performance degradation because
the partitioner is unaware of
this bandwidth difference and
weights all communication equally
as shown in fig.~\ref{fig:comm-imbal}.
Hence heavily optimised
intranode communication
actually amplifies the
problem due to
the large skewness in 
communication times.
Contributing further to
this imbalance is
the difference between
the send/receive message
sizes across partitions.
Overlapping communication
with computation using
non-blocking send/receive
can mitigate communication
imbalance. Situations
do exist where overlap of
communication and computation
is impossible
due to inherent synchronisations
in the numerical implementation
or this programming style needs extreme programming
effort~\cite{Rabenseifner2003a}.

\subsection{Redundant computations at halo entities}
\begin{figure}
    \centering
    \includegraphics{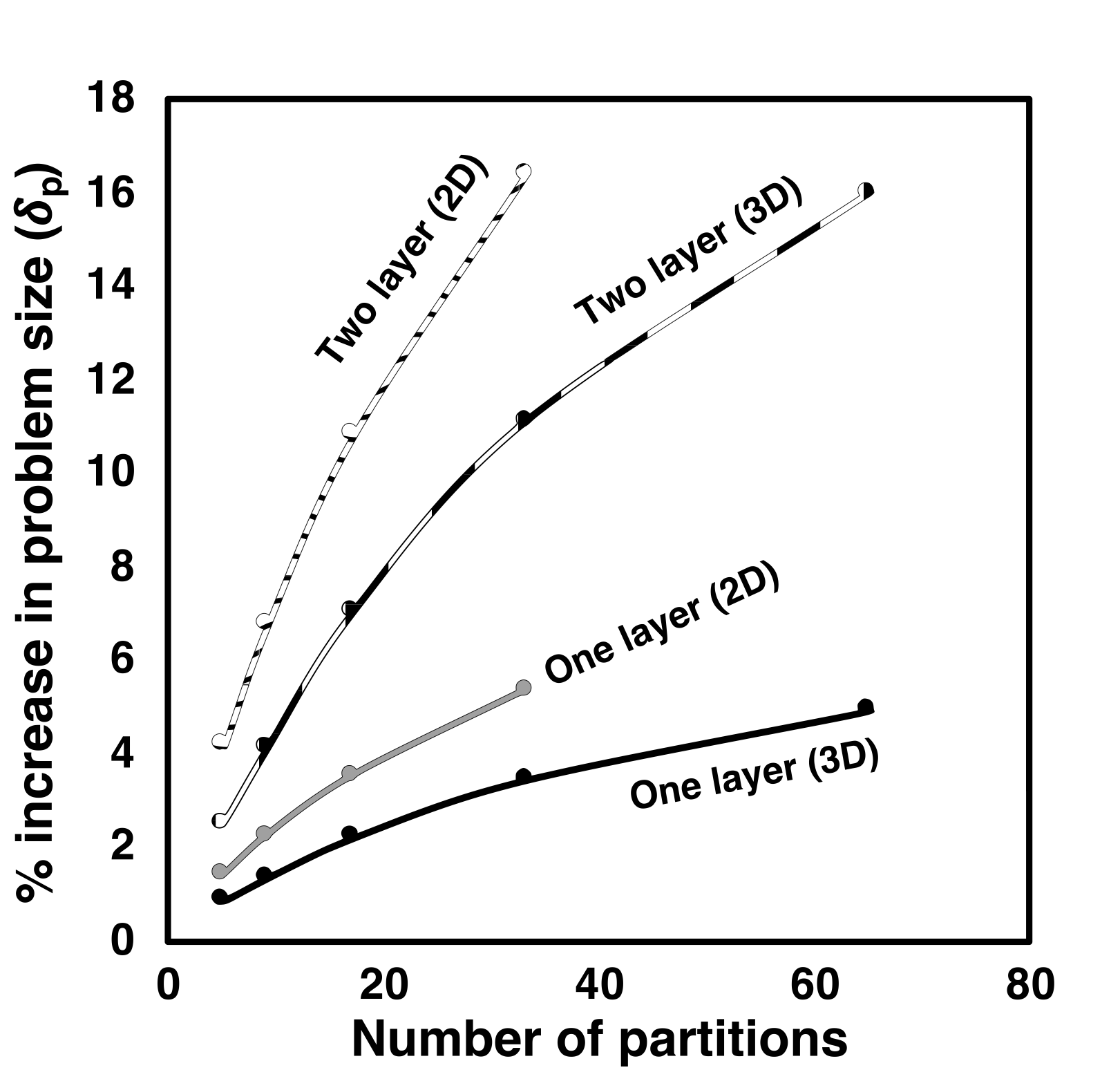}
    \caption{Increase in problem size for one and two layer overlap for 2D and 3D unstructured mesh (METIS graph partitioner)}
    \label{fig:halo-prob-size}
\end{figure}

Most scientific codes based on MPI
rely on halo entities to represent
off-processor data necessary for
local computations. Preconditioners
based on the domain decomposition
(for example additive Schwarz)
also employ large overlap
in the partitions for convergence
acceleration.
However using
overlapping (halo) entities
introduces redundant computations
resulting in an increase of
the original problem size.
Therefore
the scalability of the code deteriorates
because one now solves a larger
problem compared to the original.
In figure~\ref{fig:halo-prob-size}
the percentage increase in
problem size ($\delta_p$)
for a 3D and 2D mesh with
one layer and two layers
of overlap is shown.
With larger overlap we
find that the partition
size increases quite
rapidly (larger slope) with
processor count.

\begin{figure}
    \centering
    \includegraphics[scale=0.07]{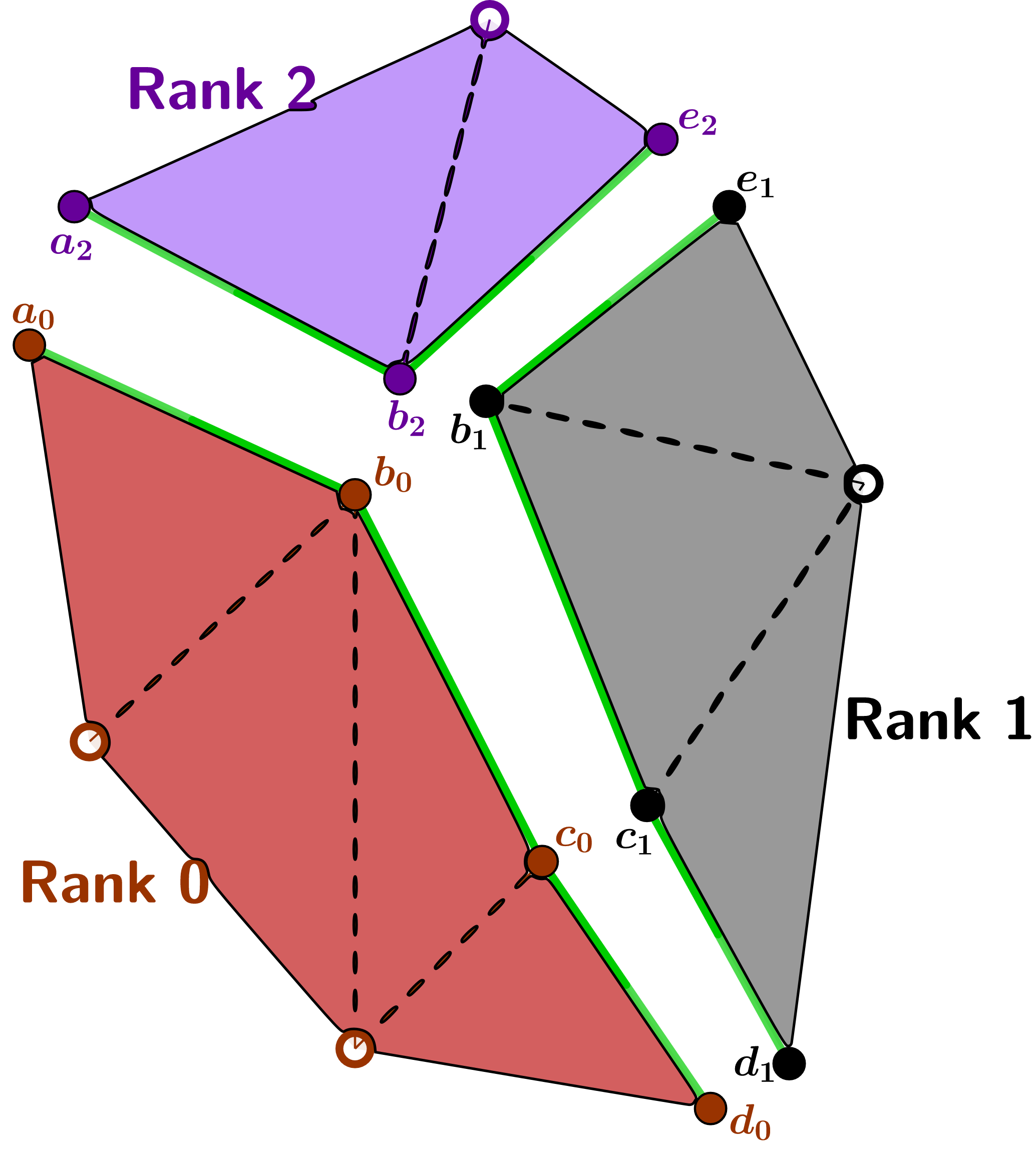}
    \caption{Shared nodes in an unstructured mesh partitioned into three ranks $0-2$; labels $a-e$ are used to denote the shared nodes}
    \label{fig:shared-nodes}
\end{figure}

Hybrid programming has the
advantage of employing
lower number of partition counts. Hence for the
same number of physical cores we considerably reduce
redundant computations.
Our in-house solver
implements a cell-vertex
finite volume scheme based
on a zero-halo partitioning
of the mesh i.e., zero
overlapping elements. The
numerical algorithm therefore
contains almost no redundant
computation making our
solver extremely
scalable and efficient.
A zero-halo partitioning
does produce redundant nodes
or shared nodes across
the partition boundary. But
they only add to the memory
overhead and not the
computation time.
We show a zero-halo
partitioned mesh
in figure~\ref{fig:shared-nodes}
along with the shared
nodes for reference.

\subsection{Memory locality}
\begin{figure}
    \centering
    \includegraphics[scale=0.2]{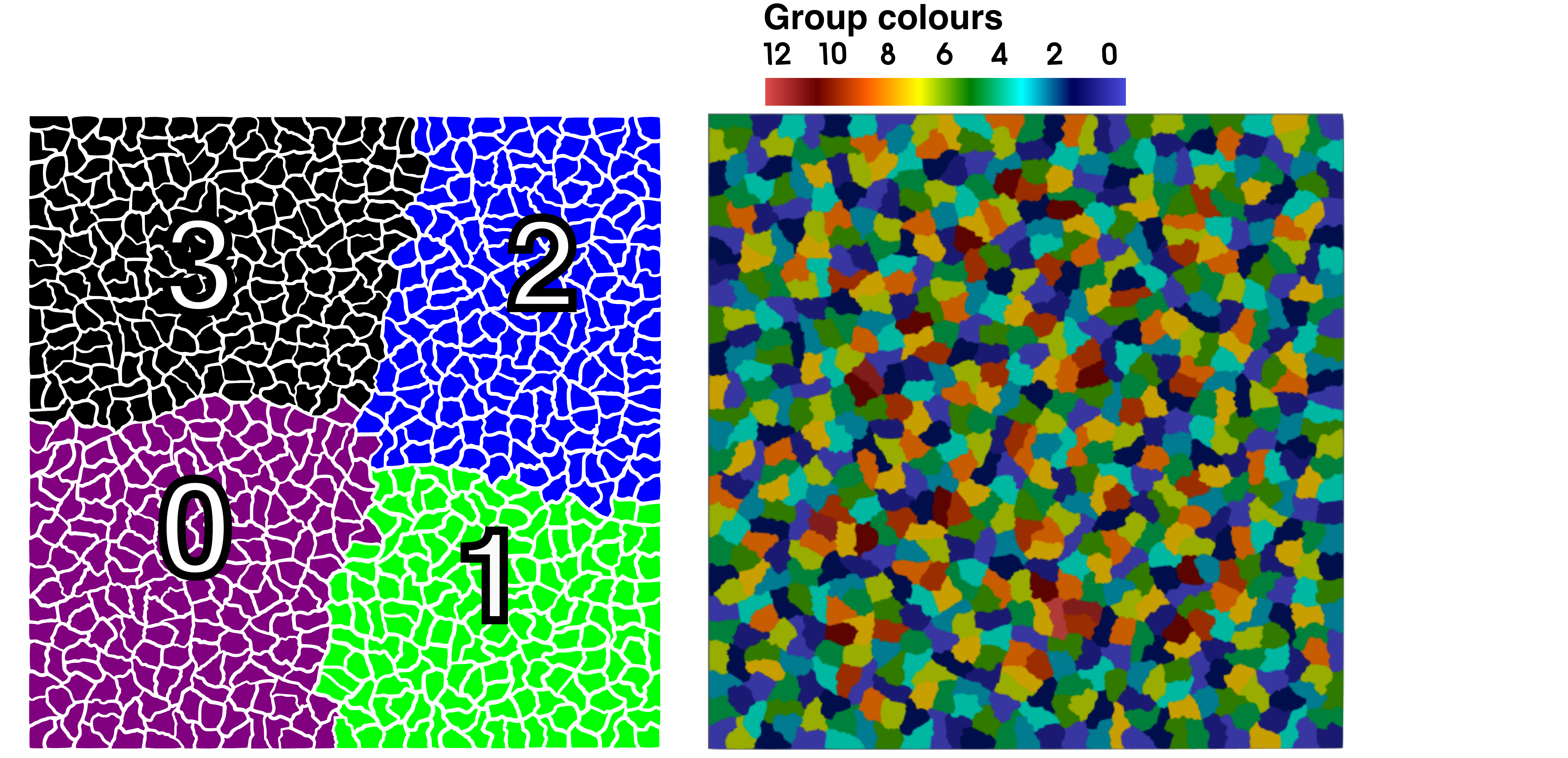}
    \caption{Cache block partitioned unstructured mesh (i) distributed across 4 MPI ranks (single node) and (ii) graph colouring for running on 4 threads (single node)}
    \label{fig:colouring_mpi_hybrid}
\end{figure}

Unstructured mesh solvers are mostly
bandwidth limited and therefore
sensitivity to memory locality.
With an increase in partition
count the data locality of
irregular applications
improve substantially
resulting in excellent
cache usage.
Sometimes they even
exhibit super-linear
scaling because the problem
fits perfectly within the
cache block~\cite{Ristov2016}.
Usually data and computation
reordering~\cite{Mellor-Crummey2001a}
is performed
to mitigate such effects.
Among such approaches
the cache block
partitioning
(i.e., dividing the
partition into
cache block sizes)
is found to be
the most
effective~\cite{Douglas2000}.
Hybrid implementations
require colouring
to avoid data dependencies
in computational loops.
Here each cache block
size partition 
belonging to the same
colour is
concurrently
run on independent
threads of execution.
In figure~\ref{fig:colouring_mpi_hybrid} 
we shown an unstructured mesh
divided into cache blocks;
(i) on the left are blocks
distributed to 4 MPI ranks
and (ii) on the right
every block is
partial 
distance-2~\cite{Gebremedhin2005} coloured
to remove data dependencies.

Clearly colouring 
destroys memory
locality due to
indirect addressing
of data inside the loop.
Lock and mutex are quite
inefficient approaches
to remove race condition
and always avoided in
computational loops.
Pure MPI codes do not
suffer from this problem
because the mesh and associated
data is contained in a given
partition making memory
access efficient. 
Virtually no
difference
in performance exists
between threads
and processes
in a modern Linux 
operating
system~\cite{Ben-Yossef2009}.
Therefore access to local variables,
memory, instructions,
etc., within a thread
or within a process
incur similar costs.

\section{Solutions to MPI problems at scale}

Overall it is
surprising to note that
switching to a
hybrid MPI 
approach
does not
solve any of the issues
outlined in the previous section
except for the absence of
internode messages.
Problems are 
merely alleviated
by reducing the
number of MPI ranks.
The issue of non-optimal
processor placement
is mostly due to the lack of an
efficient distributed graph
topology implementation in
most MPI runtime.
Therefore the choice of
MPI vs. hybrid (MPI+X)
is purely based on
load imbalance and
number of redundant
computation. 
We feel that when an
MPI code does not address
the issue of load imbalance
one cannot make a fair
comparison between
an MPI and hybrid (MPI+X)
implementation.
In fact
fixing this
issue
will equally benefit
MPI and hybrid (MPI+X)
implementations.

Using
static mesh partitioners
as load balancers for
unstructured meshes
is a major bottleneck for
MPI codes. Therefore
we found
a need for a dynamic
load balancer with built-in
hardware topology
awareness. 
Hierarchical
partitioning and
load-balancing
approach~\cite{Teresco2005b, Kong2018}
is quite
effective in solving
the major problems of
(i) optimal partition
placement, (ii)
load/communication
imbalance and
(iii) online 
partitioning at scale.
This motivated the
development of the
open-source topology aware
hierarchical unstructured
mesh partitioning tool
TreePart to address
these issues.
We found only one
open-source alternative
namely
Zoltan~\cite{Devine2002}
for such purposes.
Zoltan is now abandoned
in favour of Zoltan2;
which is still in
development and lacks
hierarchical partitioning
support.
In addition Zoltan
is not hardware aware
and
contains
MPI2 function
which are not scalable
for many MPI ranks.
For example the
migration routines
in Zoltan uses
the inefficient
\textit{MPI\_AlltoAll} and
\textit{MPI\_AlltoAllv}
functions. Also pack-unpack
functions are used in many
parts of the code which
can mostly be avoided using
MPI3 one-sided
communication.
In the next section
we provide a brief
introduction to the 
TreePart library and 
the algorithms
implemented to enable
scalable
hierarchical
mesh partitioning.

\section{TreePart Overview}

TreePart is a topology-aware hierarchical unstructured mesh partitioning and load-balancing
tool. It interfaces to existing
partitioning tools like 
Zoltan, Metis,
ParMetis, and PT-Scotch
(support for KaHIP~\cite{Schulz2017}
is under development)
and provides a scalable
infrastructure for use
in high MPI rank counts.
Presently we tested
the system up to
$26k$ MPI ranks.
The tool uses the 
hwloc~\cite{Broquedis2010}
library to query the
hardware hierarchy
and builds
internal hierarchical
MPI communicators
for each level at
run-time.
The unstructured mesh
is input as distributed
chunks which can be
partitioned
in any of the
hierarchical levels
by simply specifying one.

At present the library
takes a top-down approach
to hierarchical mesh 
partitioning~\cite{Rabenseifner2013}
i.e., the partitioning
starts at the highest
level and cascades
to the lowest
hierarchical level.
In a bottom-up 
approach~\cite{Moureau2011}
the application data is
partitioned at
the lowest hierarchical
level and cascades
up to map optimally on
to the higher hierarchies.
This results in 
a well balanced communication
load but to construct
the lowest level
sub-domains a bootstrapping
process using a
top-down partitioner
is necessary.
Ideally a 
hierarchical mesh
partitioning tool
must support both paradigms.

\subsection{Hardware topology builder}
\begin{figure}
    \centering
    \includegraphics[scale=0.35]{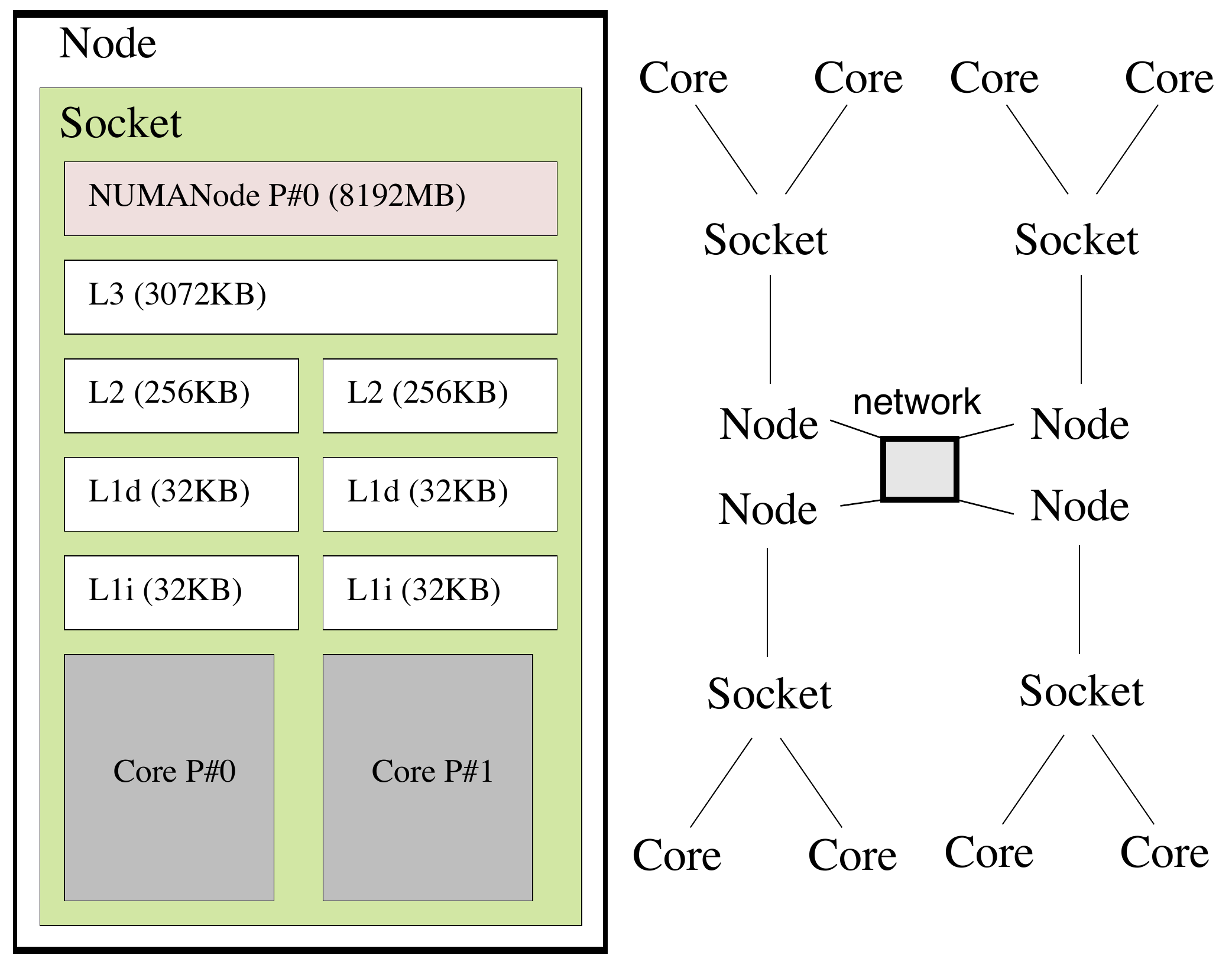}
    \caption{(i) Topology of a node with one socket and two core with shared memory and individual/shared caches (left) (ii) Topology of four nodes on a network}
    \label{fig:lstopo_1}
\end{figure}

\begin{figure}
    \centering
    \includegraphics[scale=0.3]{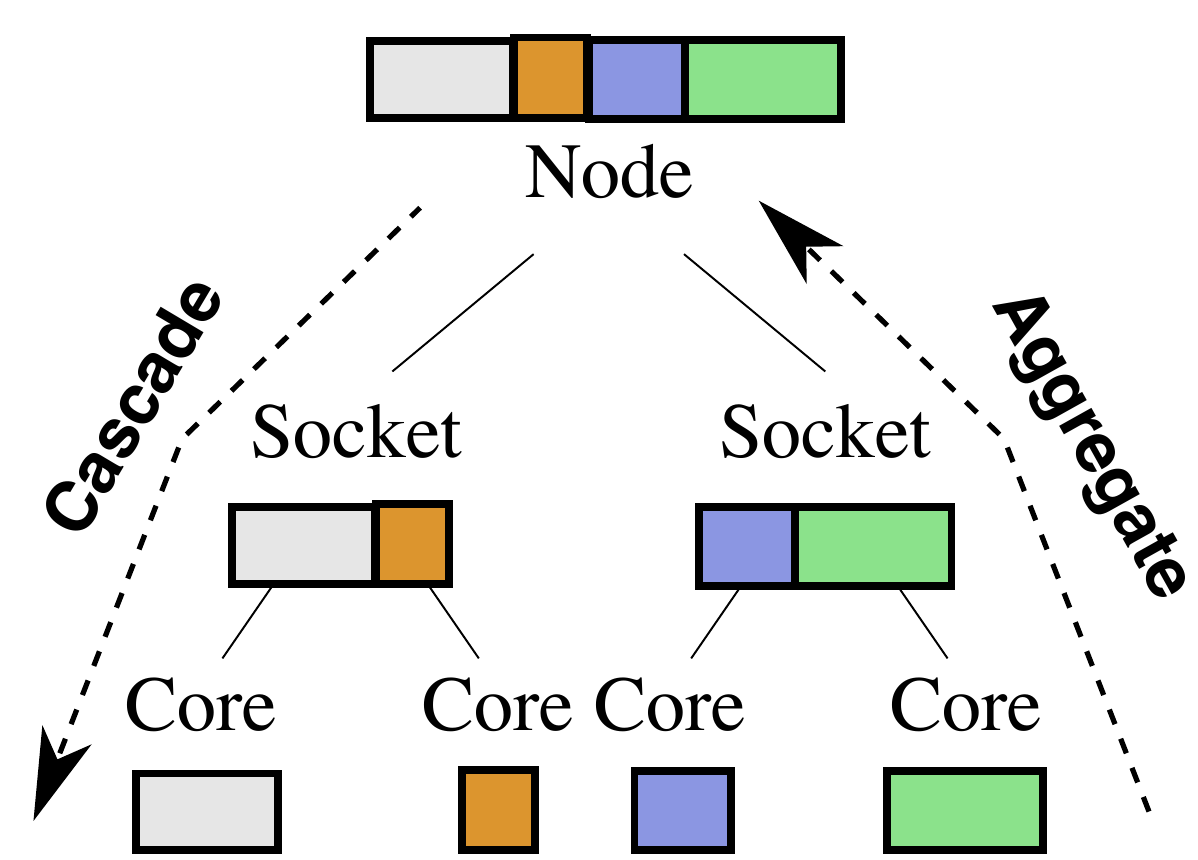}
    \caption{Aggregate and cascade operations on a node with two multi-core sockets (payload is denoted by the coloured boxes)}
    \label{fig:lstopo}
\end{figure}

To understand the structure of
the MPI tree topology built
by TreePart one must understand
few terminologies. Nodes are
the individual computing
units connected by a
common network to create a
computing pool or cluster.
Each node is made up of
multiple computing units
(CPU) called sockets
sharing a common NUMA
memory pool. Note that
accelerators like GPGPUs
can also be included as
additional sockets to
the given node.
A socket comprises of multiple
cores that can simultaneously
execute instructions. An
example architecture and
its topology is shown in
figure~\ref{fig:lstopo_1}.
For the three level
hierarchy shown in
figure~\ref{fig:lstopo}
we create three MPI
communicators, namely
node, socket, and core
communicator. Except
for the node communicator,
the socket and core
are local to a node.
% Note that TreePart
% allows modifications
% (add or remove levels)
% to the hierarchical
% structure on-the-fly
% (apart from the ones
% detected using hwloc).

Now we define two types
of hierarchical communication
called the
(i) aggregation and (ii)
cascade. In an aggregation
process,
data in a bottom level
hierarchy is aggregated
and moved to a top level
one and cascade does
just the opposite. Both
operations are highly
scalable because it
does not involve network
traffic across a node
(always within the
shared memory node).
The data used for
the aggregation or
cascade is called
the \emph{payload}.
Any operations
across a node is
only performed at
the top-most level.
This way we can
control or restrict
the number of ranks
that participate in
global communications
to avoid network traffic.
In fact nodes can be
further aggregated 
based on
their proximity to
a network switch
using the netloc
library~\cite{Goglin2015}
but we post-pone
this study as a future
extension to TreePart.

\subsection{One-sided distributed dictionaries}
An integral part of TreePart
is the one-sided distributed
dictionary (ODD) data-structure.
ODD is used to store unstructured
mesh entities like nodes, elements,
boundary faces, etc., as 
key-value pairs and
perform scalable range
queries on keys.
The directory consists of
arrays of key-value pairs
distributed across any one
of the hierarchical levels.
The keys are of integral types
and the values can be
plain-old-data (POD) or
variable length vectors (unlike
ZoltanDD~\cite{Devine2002}).
This
assumption
enables us to store
the data already in a
packed format in the
directory structure for
transmission.
We use the assumed
partitioning~\cite{Falgout2006}
method of Falgout et al. to
initially distribute the
keys globally and to
perform scalable range
queries on keys. During
the rendezvous phase
to determine the neighbours
we use a blind send (or
receive) query using one-sided
MPI primitives, which is
explained in detail in the
following subsection.
In current version of
TreePart we assume that
the level which initiates
the queries to the
ODD is at the same
level as the ODD itself.
Therefore the user is responsible
for aggregating/cascading
queries from a different
level. But we plan
to remove such restrictions
in a future version.

\subsection{Scalable blind send (or receive) queries}
The blind send (or receive)
problem can be described
as follows. A set of
MPI ranks called the
\emph{blind initiator}
intends to
send (or receive)
data to (or from) its
neighbouring MPI
rank. The 
neighbouring
rank (referred
to as the \emph{blind end})
is unaware of the
fact that it must
cooperate with the
receiving (or sending)
act.
Usually an all-to-all
type MPI primitive
is used to
transpose the
send (or receive)
matrix and obtain
this information. 
We propose
a scalable
approach using
one-sided accumulate
primitive. Firstly
we create an MPI
window to expose
an increment variable
initialised to zero.
Then the blind
initiator calls 
an one-sided
accumulate to its
blind end variable
which increments
the value by one.
Once all blind
initiator have completed
the accumulation; the
blind end will have
the number of ranks
it is expected to
cooperate with. A
pseudo-code in C++
is provided in 
listing~\ref{lst:count_blind}.
\begin{lstlisting}[language=C++, caption={Counting blind send (or receive) C++ pseudo-code}, label={lst:count_blind}]
int BlindCount(list<int> &rank_list) {
 int count = 0;
 int increment = 1;
 MPI_Win count_win;
 MPI_Win_create(&count, sizeof(int), 
   1, MPI_INFO_NULL, MPI_COMM_WORLD,
   &count_win);
 MPI_Win_fence(0, count_win);
 for (auto rank : rank_list)
  MPI_Accumulate(&increment, 1, MPI_INT,
    rank, 0, 1, MPI_INT, MPI_SUM, count_win);
 MPI_Win_fence(0, count_win);
 MPI_Win_free(&count_win);
 return count;
}
\end{lstlisting}

This minimal information
is sufficient to
receive the data
from all blind
initiator of a
blind end by
probing for the message
size using 
MPI\_ANY\_SOURCE (and
MPI\_ANY\_TAG),
allocating the buffers
and finally completing
the send or receive request.
A brief pseudo-code in C++
is provided in
listing~\ref{lst:blind_comm}.
\begin{lstlisting}[language=C++, caption={Blind communication (C++ pseudo-code)}, label={lst:blind_comm}]
for (int i = 0; i < count; ++i) {
 MPI_Probe(MPI_ANY_SOURCE, MPI_ANY_TAG,
   MPI_COMM_WORLD, &stat);
 MPI_Get_count(&stat, MPI_BYTE, &size_in_bytes);
 // Allocate buffer using size
 ...
 // Do a receive or send
 MPI_Recv(&buffer, size_in_bytes, MPI_BYTE,
  stat.MPI_SOURCE, stat.MPI_TAG, MPI_COMM_WORLD, &stat);
}
\end{lstlisting}
The blind send combined with
the assumed partition
algorithm is a highly 
scalable infrastructure
for ODD queries and
we fully avoid the use of
the inefficient
all-to-all type primitives.
In addition, we use ODD and
blind communication to
avoid \textit{Allgather}
type MPI
operations to construct
the dual graph of the
mesh necessary for
partitioning the
mesh cells using
graph partitioning
libraries.

\subsection{Hierarchical partitioning}
\begin{figure}
    \centering
    \includegraphics[scale=0.07]{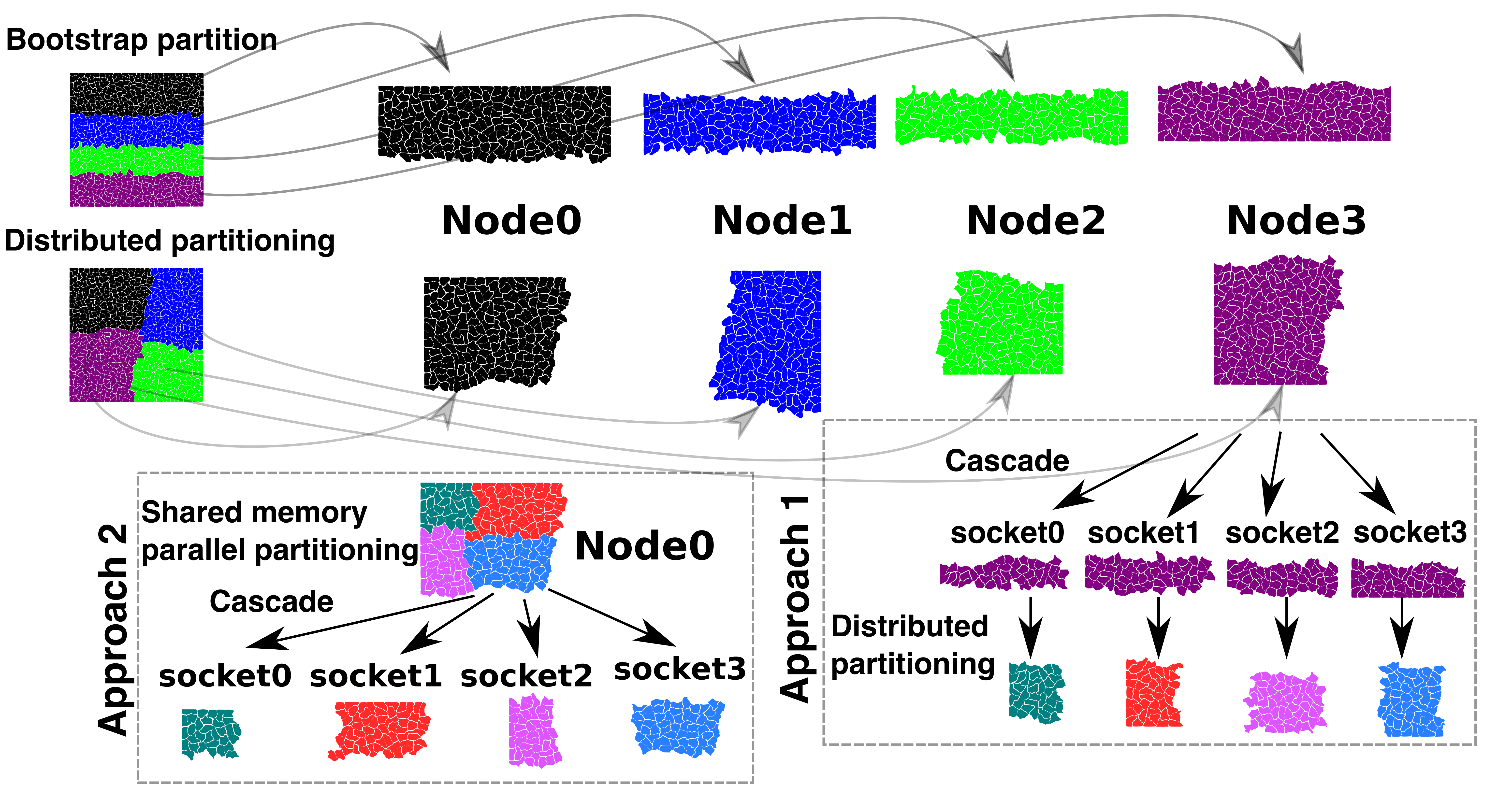}
    \caption{Bootstrap partition and approaches (1) and (2) to hierarchical partitioning considering two level node-socket hierarchy}
    \label{fig:bootstrap}
\end{figure}
TreePart currently implements
a top-down hierarchical
partitioning approach.
The partitioning therefore
starts at the top most level
of the topology referred to
as the \emph{bootstrap 
partition level} (BPL).
To start the process
at BPL one first
requires an approximately
partitioned mesh
available at this level
(see figure~\ref{fig:bootstrap}).
The payload can
be positioned in any
arbitrary level
(usually the bottom-most
level)
and one has to
aggregate the payload
to the BPL. Then
based on the users
input the particular
partitioning scheme
is chosen and the
mesh is redistributed
based on the new partitioning.
An illustration
of this process
is shown at the
top of 
figure~\ref{fig:bootstrap}.
Once the BPL partitions
are obtained there are
two ways the partitioning
can proceed, (1) cascade
equal mesh chunks to the
next level and
perform distributed parallel
partitioning or
(2) perform a shared-memory
parallel partitioning
in the current level
and cascade partitions
to the next level. 
An illustration of both
approaches is shown
in 
figure~\ref{fig:bootstrap}.
Approach
2 is attractive because
we completely avoid
MPI communication and
it also blends well with
hybrid programming models.
The partitioner terminates
upon reaching
the last level.

We provide two ways to
interface the
unstructured mesh data to
TreePart.
In the library approach
the mesh is assumed to be
already divided (although
not optimally) among all
the participating ranks
(lowest level) and
supplied to TreePart 
using a ParMETIS style
API function call to
generate a refined partition.
The mesh data is aggregated
to BPL to start the
partitioning process.
In the second approach
we provide a standalone
tool which can read
two popular formats
CGNS unstructured mesh
(experimental) and
our in-house
HiP HDF5 unstructured 
mesh format. The user
executes the tool
to generate HDF5 partition
mesh files that can 
be read later for further
processing.

\subsection{ShHalo framework for shared-memory halo communication}
\begin{figure}
    \centering
    \includegraphics[scale=0.5]{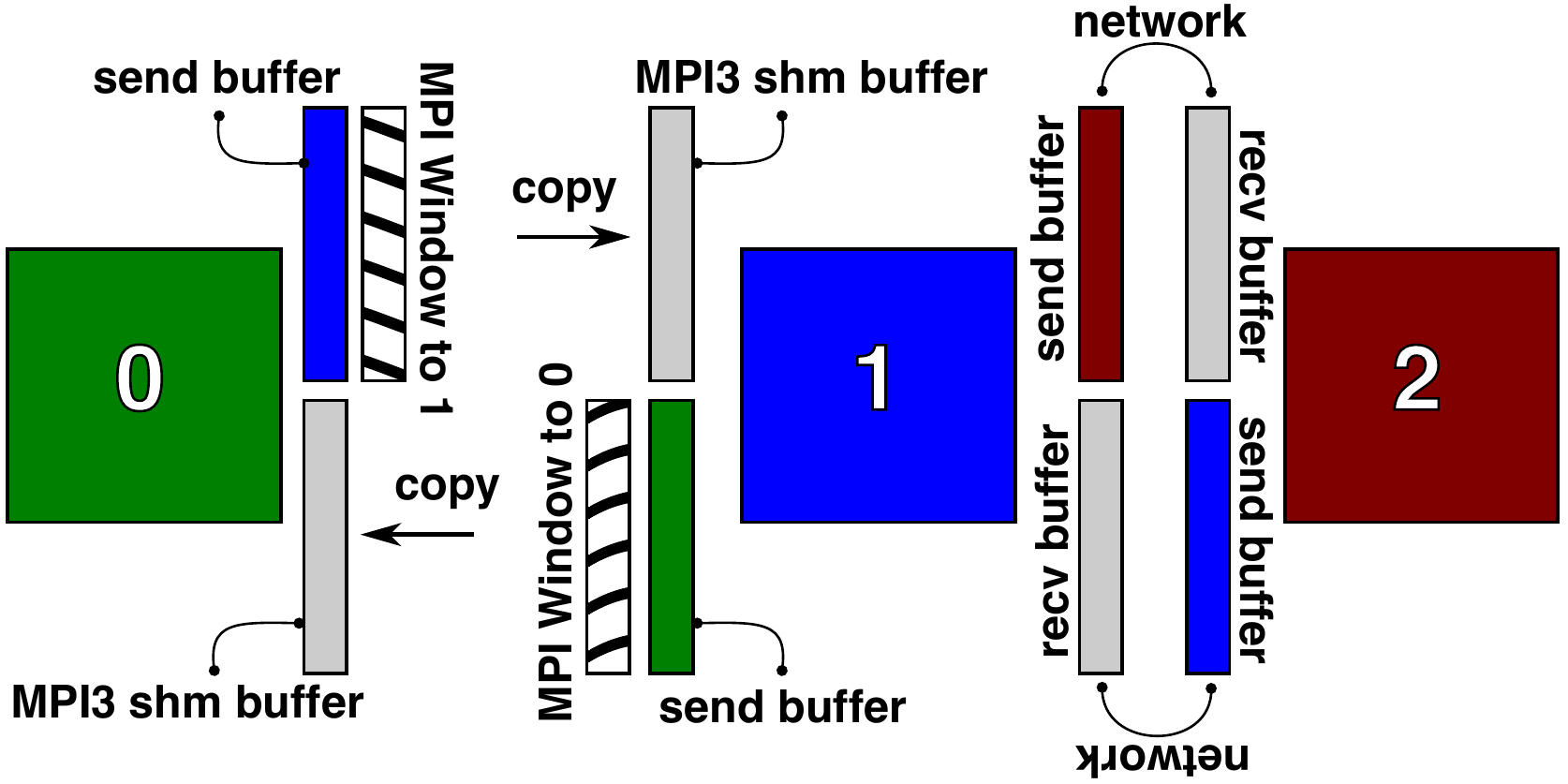}
    \caption{MPI3 shared-memory optimisation in ShHalo for intranode rank $0-1$ and normal network send/receive for internode ranks $1-2$}
    \label{fig:shhalo}
\end{figure}

MPI implementation
that do not support
kernel direct
copy result
in inefficient
intranode
communication.
We encounter such
situations quite often
in practice.
To
handle such situations
TreePart provides
ShHalo, which
is an
MPI3 shared-memory
optimised halo
communication
module. ShHalo
wraps the original
MPI calls for
halo data
communication and
segregates them
into inter and
intranode communicators.
The internode
requests are
initiated using
network send and
receive primitives.
For intranode
requests the data is
copied directly
to the neighbouring
processor receive halo
buffers
using MPI3 
shared-memory
primitives
as shown in
figure~\ref{fig:shhalo}.
The users has
to provide the halo
communication
schedule to
aid ShHalo to
pre-allocate
the shared-memory
buffers. Note that
MPI3 shared-memory
requires the use
of its own allocation
routines. Therefore
one extra receive
buffer is allocated
for every
halo entity. This
approach requires
minimal
modification to
the existing
halo transfer
in order to
use ShHalo.
A more intrusive
approach is to
get rid
of halo 
intranode
halo buffers
in the users code
and directly expose
the neighbouring
intranode
processor's memory.
So the intranode
transfer becomes
simple direct
memory-to-memory
copy of halo data
without any
intermediate
buffers
(zero-copy
)~\cite{Besnard2015a}.

\subsection{Hierarchical load-balancing}
\begin{figure}
    \centering
    \includegraphics[scale=0.32]{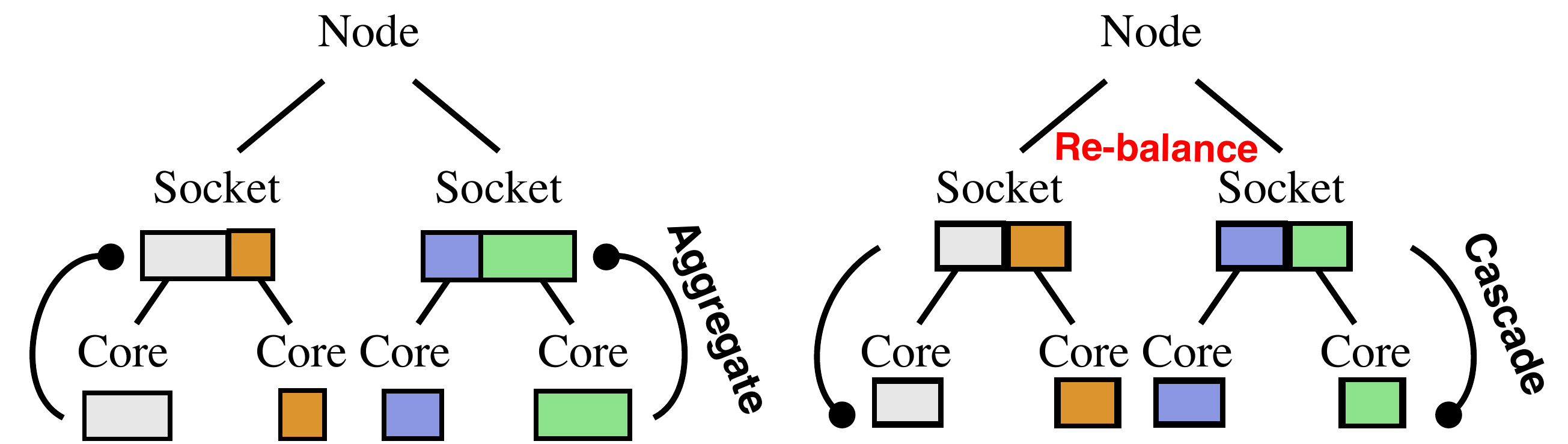}
    \caption{Re-balancing at a shared memory hierarchical level is three step process,  aggregation, re-balance, and cascade}
    \label{fig:hier-bal-shm}
\end{figure}

Hierarchical load-balancing
is the most distinguishing
feature of TreePart. We have
implemented mesh re-balancing
using nodal and element weights
that can work at any given
hierarchical level. This
gives the user the advantage
of using the fast and scalable
partitioning at the
lowest level more frequently.
TreePart uses shared-memory
parallel partitioning
at lower levels thus
totally avoiding MPI
communications
as shown in
figure~\ref{fig:hier-bal-shm}.
The re-balancing
cost is drastically
reduced since we
restrict the movement
of data within
the node.
The expensive
node level re-balancing
of the mesh can be
called less frequently
to considerably reduce
the re-balancing
cost.
The user can optimise the
re-balancing frequency across
levels for his 
application
to achieve significant
reductions in run-time.

\section{Results}
\begin{figure}
    \centering
    \includegraphics[scale=0.4]{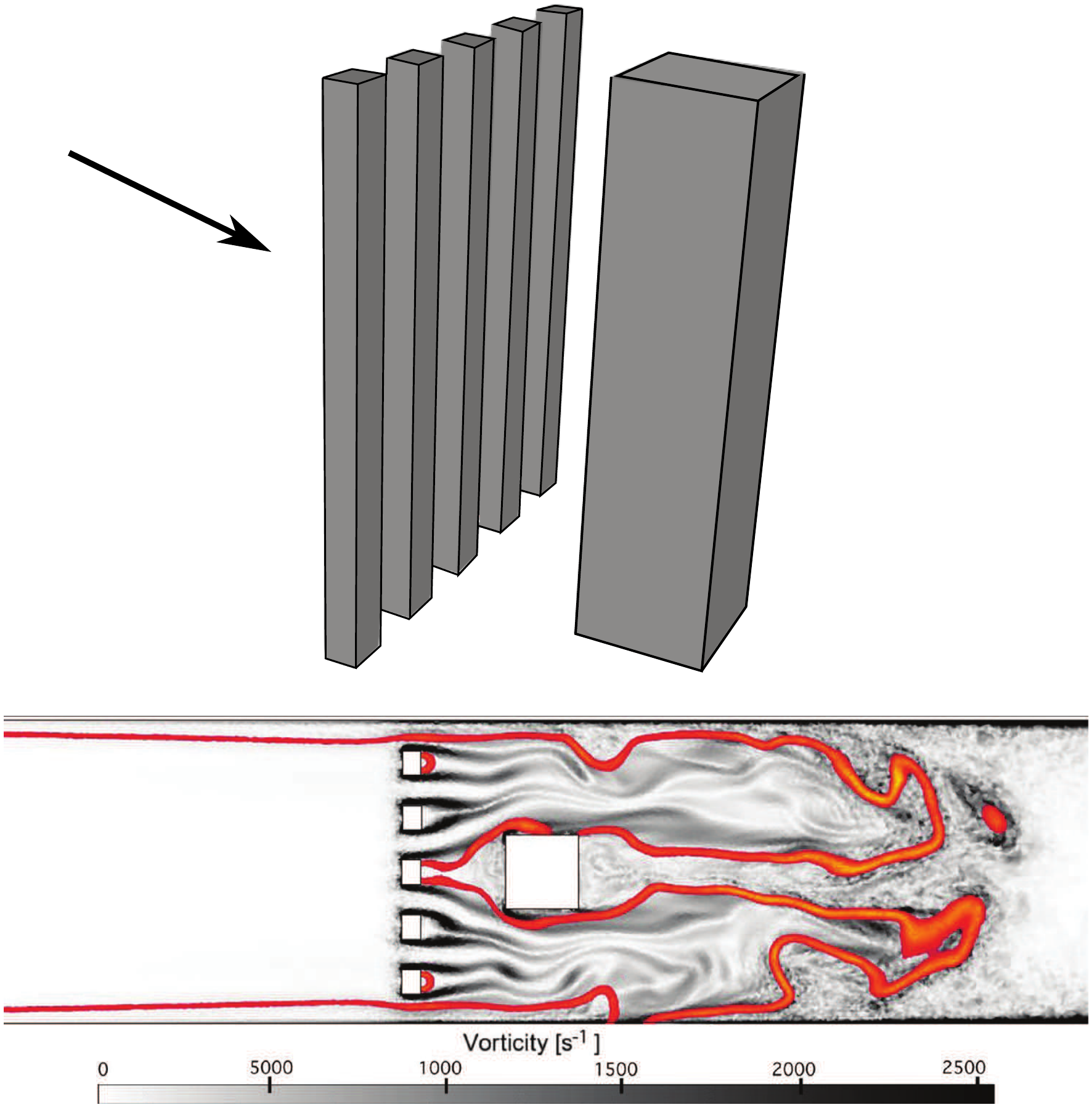}
    \caption{LES of flame propagation;
    solution plotted at time instance $13.5$~ms (Vorticity field and isoline of heat release); Geometry of the vertical obstacles shown on top.}
    \label{fig:perf150M}
\end{figure}

We interfaced TreePart
library with
our in-house solver AVBP
using the ParMETIS
style API of TreePart.
We study the deflagrating 
flames of compressed
natural gas
(CNG)~\cite{Masri2012} using
large eddy simulation
(LES) with a two step
reaction model.
The computational
domain is an
unstructured mesh
with $150$M
tetrahedral elements.
The transient,
turbulent flame fronts
that propagate past a
sequence of solid 
obstacles, whose
geometry and
computational
domain is shown in
figure~\ref{fig:perf150M}.
The chamber is square
in cross-section with internal dimensions of length, $250$~mm and side, $50$~mm producing an overall
volume of $0.625$~l
and a length to width ratio
of 5. The arrow in
figure~\ref{fig:perf150M}
indicates the direction
of the flame propagation.

\subsection{Solver scalability with  hierarchical partitioning}
\begin{figure}
    \centering
    \includegraphics[scale=0.6]{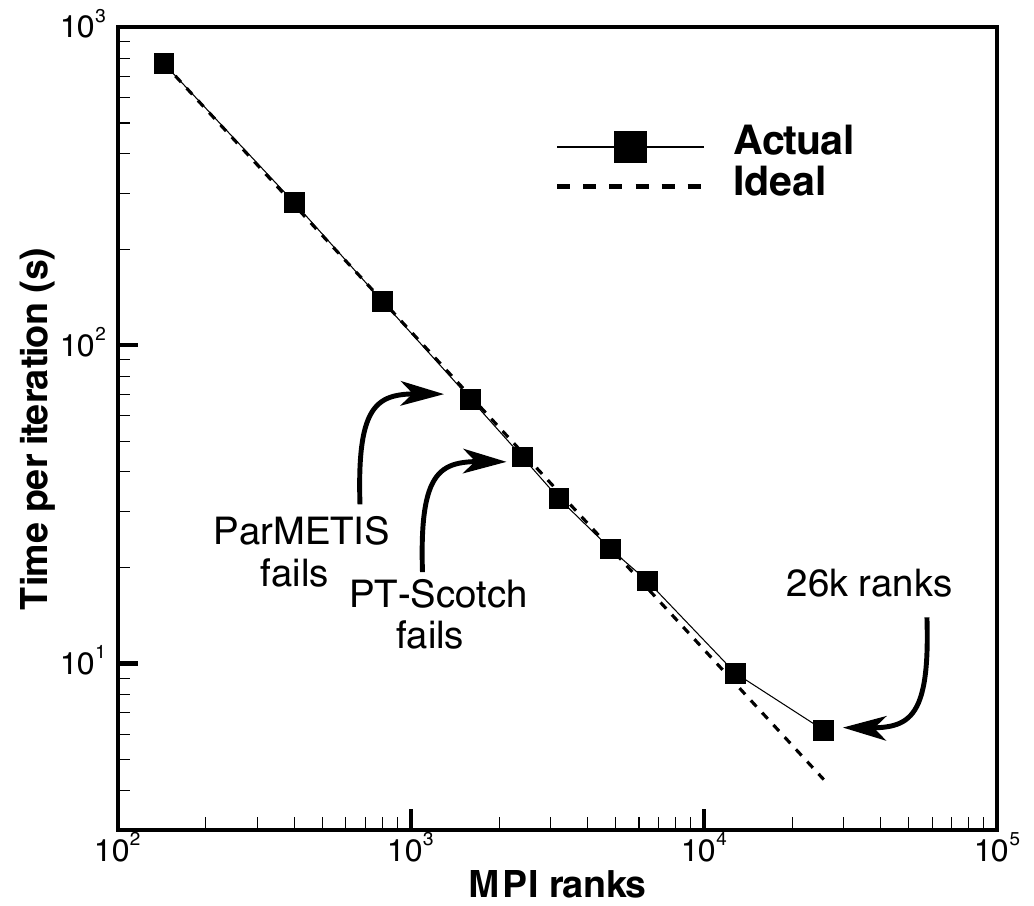}
    \caption{Strong scaling of combustion test case using  TreePart (online)}
    \label{fig:weak-scale}
\end{figure}

The strong scalability results
of the combustion test
case using online
hierarchical partitioning
for different MPI rank
counts (=\# of partitions)
is shown in 
figure~\ref{fig:weak-scale}.
Note that ParMETIS failed
to generate online partitions
for this test case beyond
$2k$ MPI ranks and 
PT-SCOTCH failed beyond
$5k$ but the hierarchical
partitioner
successfully generated
partitions online
beyond $25k$ MPI ranks.
The plain graph
partitioners failed
due to MPI exit errors
in the all-to-all function.

\subsection{Effect of processor placement and intranode communication}
Hierarchical partitioning
places MPI ranks
efficiently on to
processor
cores and avoids
unnecessary inter
node communication.
In addition ShHalo
MPI3 shared memory
optimisation can
be applied to further
optimise intranode
halo data communication.
To show the effect
of processor placement
and the ShHalo optimisation
we ran four test cases with
different partitioning scheme,
(1) plain non-hierarchical
parallel graph partitioning,
(2) plain non-hierarchical
geometric recursive coordinate
bisection,
(3) topology-aware 
hierarchical geometric
recursive coordinate bisection
(RCB) without ShHalo, and
(4) topology-aware
hierarchical RCM 
with ShHalo
optimisation. Note
that the RCB partitioner
gives lower quality
partitioning compared
to a graph based
partitioner because
it does not consider
edge-cut information.
On the contrary it
produces almost perfectly
balanced partitions
%with lower running
%time 
compared to graph.
We ran the cases on
10 nodes, where each
node consists of 
two sockets of 18 core 
Intel Xeon processor.
This brings the total
core count to 360.
\begin{figure}
    \centering
    \includegraphics[scale=0.5]{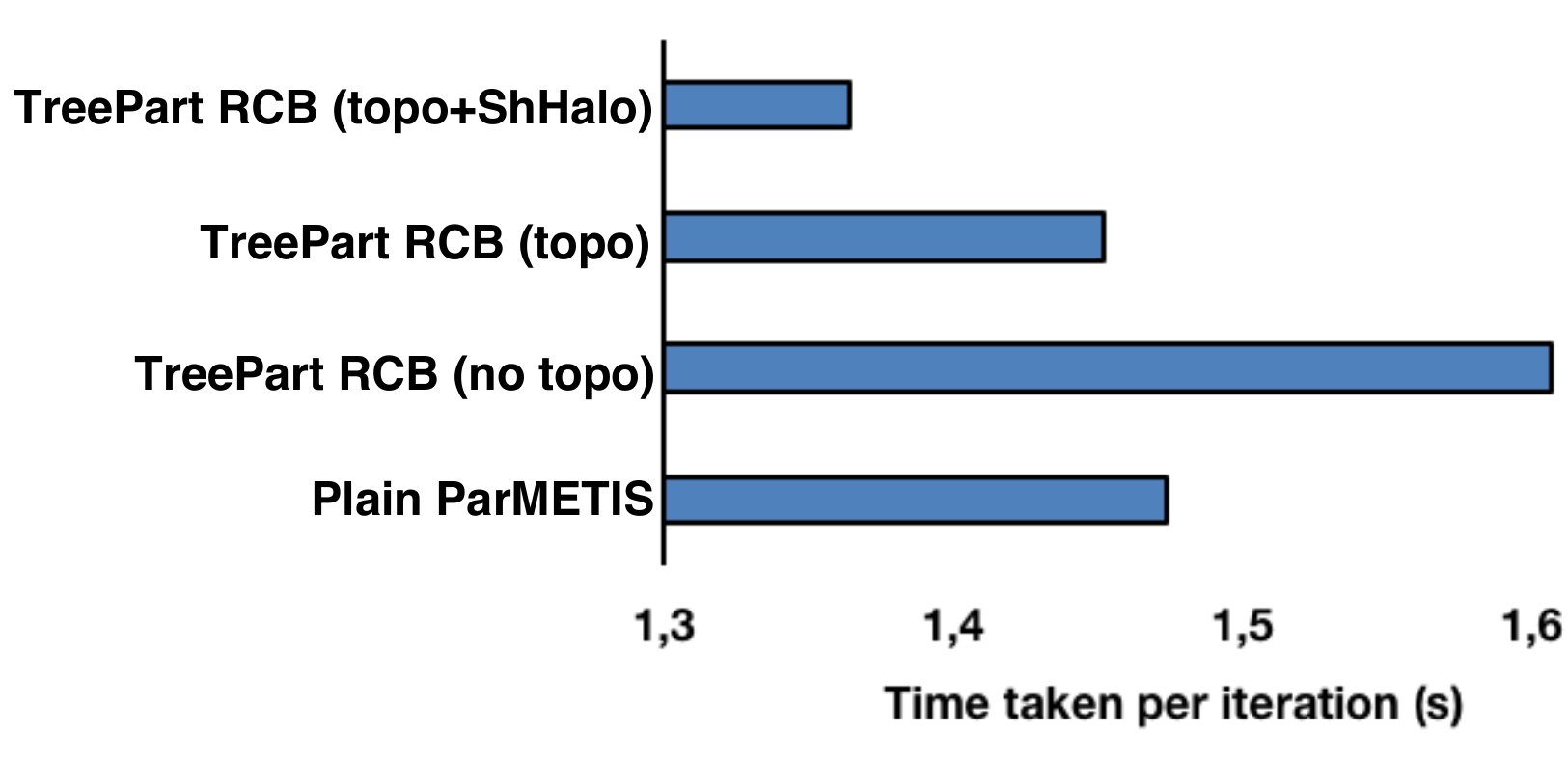}
    \caption{Comparison of solver time per iteration for the indicated partitioning strategy}
    \label{fig:hier-shhalo}
\end{figure}

The solver time taken per
iteration is plotted
in figure~\ref{fig:hier-shhalo}
for the four test cases.
As mentioned previously
the plain RCB partitioner
results are worse than
the plain graph partitioner.
We see that by placing
the partitions correctly
using topology-awareness
RCB is more efficient
than the plain graph
partitioner. With
the shared-memory
optimisation we see
even further speed up.

\subsection{Effect of hierarchical load-balancing}
\begin{figure}
	\centering
	\includegraphics[scale=0.5]{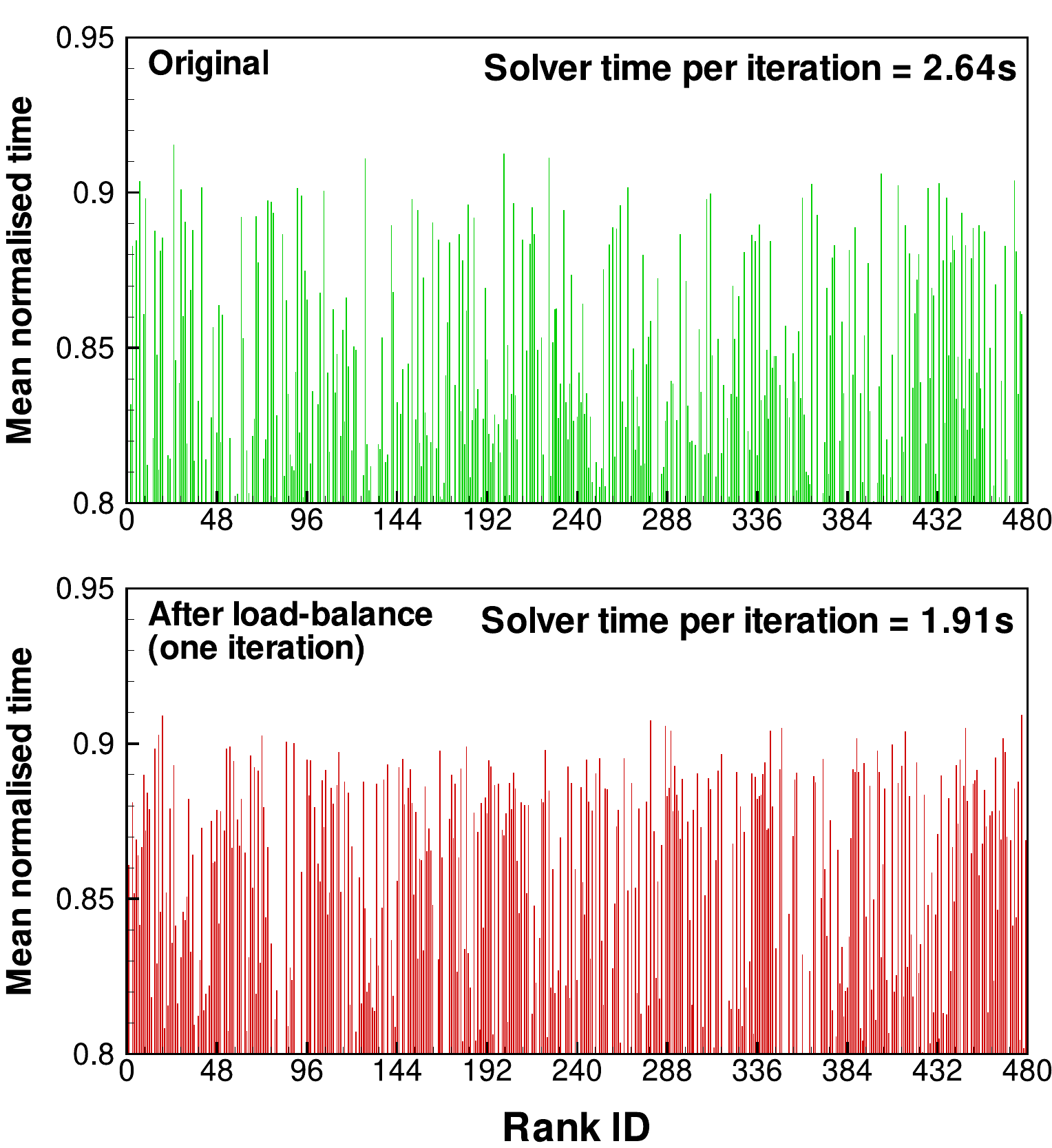}
	\caption{Mean normalised time of each partition before and after load-balancing considering $2$ sockets of $24$ cores across the $10$ bootstrap nodes each with $2$ sockets (level 1) of $24$ cores (level 2) i.e., $10 
		\times 2 
		\times 24$ hierarchy)}
	\label{fig:lb-hist}
\end{figure}

To load-balance the mesh we
require to calculate weights
of the mesh entities based on
runtime timing measurements.
The measurement must
not be too coarse grained
otherwise we risk averaging
out important deviations
in runtime that requires
balancing. Our solver
already employs an optimisation
to improve cache utilisation
by dividing mesh elements
into cache block groups
(see 
figure~\ref{fig:colouring_mpi_hybrid}).
Therefore we measured the
computational time
of each cache block and
spread this value to
the elements in the
block by dividing by the
total number of 
elements in the block.
This way we minimise the
risk of using a very coarse
grain timing and at the same
time avoid the risk of
getting erroneous
measurements by going
too fine grain.
The improvement in load
balance before and after
performing one iteration
of TreePart hierarchical
load-balancing at the
two bottom-most levels
(2 sockets and 24 cores)
is shown
in figure~\ref{fig:lb-hist}.
The case was run on 10
nodes with 2 sockets
having 24 cores each.
Even with a coarse grain
timing information and
load-balancing across
the lowest two levels 
considerable
improvement in load balance
is achieved and the
reduction in computational
time per solver iteration
is shown in
figure~\ref{fig:lb-hist}.

\subsection{Comparison of load-balancing timing across hierarchy}
\begin{figure}
    \centering
    \includegraphics[scale=0.5]{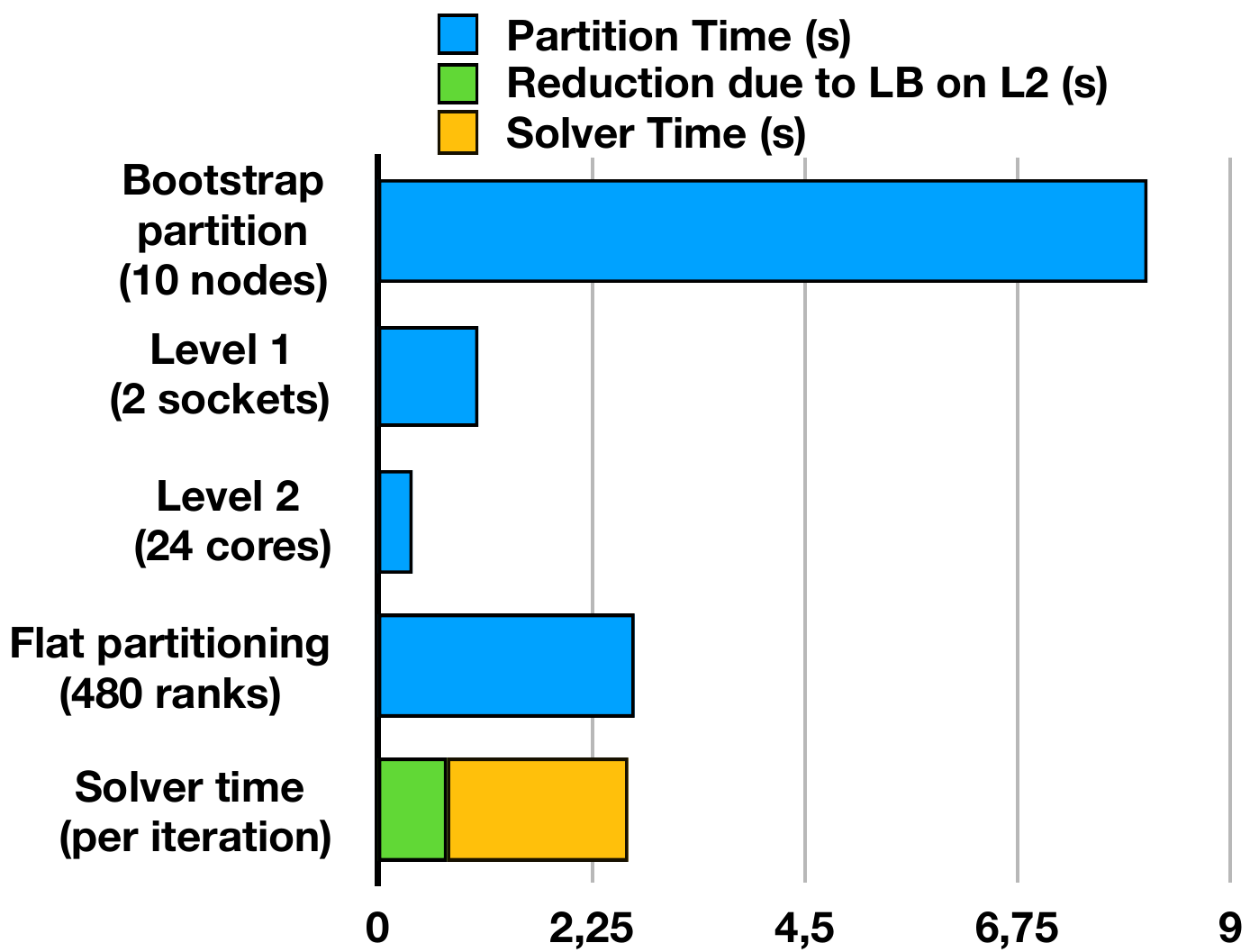}
    \caption{Comparison of load-balancing time across hierarchy of $10$ bootstrap nodes each with $2$ sockets (level 1) and $24$ cores (level 2) i.e., $10 
    \times 2 
    \times 24$ hierarchy)}
    \label{fig:lb-time}
\end{figure}

The cost of load-balancing
should be as low as
possible so that 
it does not shadow
the actual gain in computational
speedup as a result of
the load-balancing.
A hierarchical load-balancing
algorithm enables one to
load-balance
fine grain partitions at the
lowest level more frequently
than higher levels. Thus
providing more flexibility
in optimising the load-balancing
frequency. In figure~\ref{fig:lb-time}
we compare the repartitioning time
between different hierarchical
levels, flat partitioning considering
all 480 ranks and gain in solver
time per iteration due to
partitioning across shared
memory levels (1 and 2).
Firstly we notice that the
bootstrap partitioning
time is much higher than
the cost of a flat
partitioning considering no
hierarchy. This is mainly
because the RCB used is a
pure MPI implementation
and does not exploit the
shared memory cores available
at the bootstrap level.
Using a hybrid parallel RCB
algorithm will considerably
bring down the repartitioning
cost at the bootstrap level.
This also
provides an interesting use
case for hybrid parallel
graph partitioning
algorithms.
The cost of repartitioning 
at levels L1 and L2 are
considerably lower than
the flat partition. Moreover
repartitioning at level 1 or 2
does not involve communication
across the network and
highly local. The gain
in speed-up due to
repartitioning at level 1
and 2 is easily amortized
by the gain in computational
time gained per iteration.
The focus of the
present work is the
demonstration of TreePart
and its hierarchical
load-balancing ability.
Therefore we postpone the study
on the actual optimisation
of the solver and TreePart
hierarchical load-balancing
to as an extension 
of the present work.

\section{Future Work}
Parallel distributed
breadth first search
algorithm
(BFS)~\cite{Buluc2011}
is useful to
find connected
components in
a graph.
Such a BFS implementation
in TreePart can be used
to ensure connected
partitions because
distributed graph
partitions can produce
disconnected ones.
Disconnected partitions
are undesirable for
test cases involving
Lagrangian particle
dynamics, parallel mesh
adaptation, etc.
The BFS will be used to
merge the disconnected
mesh entities by re-migrating
to the correct connected rank.
We are making TreePart
more HPC aware
by interfacing with
netloc~\cite{Goglin2015}
library to build
network awareness;
map ranks optimally
by exploiting network
proximity information.
Currently the hierarchical
framework supports only
single weights. Implementing
multiple weights for 
load-balancing is planned.
Hierarchical techniques in
TreePart are quite attractive
to parallel mesh adaptation
implementation, which we
plan as a future extension.
Finally we plan to release
TreePart as open-source
software for benefit
of the HPC community.
% Currently we are extending
% TreePart by implementing
% hierarchical parallel
% mesh adaptation
% using the
% MMG3D~\cite{dobrzynski2012}
% library. In addition

\section{Conclusion}
With the advent of
advanced features
and optimisations like
MPI3 shared-memory, 
HPC awareness, etc., the
performance issues
in plain MPI codes
even for very large
number of ranks
can be made 
competitive to
hybrid MPI implementations.
This approach 
has an advantage
of introducing minimal
changes to HPC codes
to achieve the said
performance.
The critical issues
that we identified
that hinder the performance
of plain MPI codes are
(i) load balancing,
(ii) minimising redundant
computation
at halo entities
and (iii) reliable partitioning
of meshes into extremely
high partition counts
in HPC codes.
We have developed
the TreePart library
to address these
problems using a
variety of optimisations
resting on the core
principle of hardware-awareness
and scalability.
We showed excellent
improvement in our
industry strength
and highly optimised
in-house combustion LES
code using TreePart.
We hope that plain MPI
HPC codes using
TreePart is a viable
alternative and also a
companion to hybrid
MPI counterparts.

\begin{acks}
The research leading to these results has received funding from the European Union’s Horizon 2020 research and innovation programme under the EPEEC project, grant agreement No 801051.
This work was granted access to the HPC resources of IDRIS under an allocation by GENCI for the Grand Challenges Jean Zay (2019).
\end{acks}

\bibliographystyle{ACM-Reference-Format}
\bibliography{sample-base}

\end{document}